\documentclass[runningheads,a4paper]{llncs}

\usepackage{amssymb}
\usepackage{graphicx}

\usepackage{url}
\urldef{\mailsa}\path|{m.menzel, m.axer}@fz-juelich.de|
\urldef{\mailsb}\path|h.a.de.raedt@rug.nl|    
\urldef{\mailsc}\path|k.michielsen@fz-juelich.de|
\newcommand{\keywords}[1]{\par\addvspace\baselineskip
\noindent\keywordname\enspace\ignorespaces#1}

\usepackage{amsmath}
\usepackage{upgreek}
\usepackage{marvosym}
\newcommand{\envelope}{\raisebox{4pt}{\scalebox{0.9}{\Letter}}\kern-1.7pt}
\usepackage{hyperref}

\begin{document}

\mainmatter  
\title{Finite-Difference Time-Domain Simulation for\\Three-dimensional Polarized Light Imaging}
\titlerunning{FDTD Simulation for 3D Polarized Light Imaging}
\author{Miriam Menzel$^{1(}$\envelope$^)$
\and Markus Axer$^{1}$\and Hans De Raedt$^{2}$\and Kristel Michielsen$^{3}$}
\authorrunning{M.\ Menzel et al.}
\institute{$^{1 }$Institute of Neuroscience and Medicine (INM-1), \\Forschungszentrum J{\"u}lich, 52425 J{\"u}lich, Germany\\
\mailsa\\
$^{2 }$Zernike Institute for Advanced Materials, \\University of Groningen, 9747 AG Groningen, the Netherlands\\
\mailsb\\
$^{3 }$J{\"u}lich Supercomputing Centre (JSC), \\Forschungszentrum J{\"u}lich, 52425 J{\"u}lich, Germany\\
\mailsc}
\toctitle{Simulation and Modeling for Three-dimensional Polarized Light Imaging}
\tocauthor{Miriam Menzel, Kristel Michielsen, Hans De Raedt, and Markus Axer}
\maketitle


\begin{abstract}
Three-dimensional Polarized Light Imaging (3D-PLI) is a promising technique to reconstruct the nerve fiber architecture of human post-mortem brains from birefringence measurements of histological brain sections with micrometer resolution.
To better understand how the reconstructed fiber orientations are related to the underlying fiber structure, numerical simulations are employed.
Here, we present two complementary simulation approaches that reproduce the entire 3D-PLI analysis: First, we give a short review on a simulation approach that uses the Jones matrix calculus to model the birefringent myelin sheaths. Afterwards, we introduce a more sophisticated simulation tool: a 3D Maxwell solver based on a Finite-Difference Time-Domain algorithm that simulates the propagation of the electromagnetic light wave through the brain tissue. 
We demonstrate that the Maxwell solver is a valuable tool to better understand the interaction of polarized light with brain tissue and to enhance the accuracy of the fiber orientations extracted by 3D-PLI.
\keywords{Polarized light imaging $\cdot$ Nerve fiber architecture $\cdot$ Optics $\cdot$ Birefringence $\cdot$ Jones matrix calculus $\cdot$ Maxwell solver $\cdot$ Finite-Difference Time-Domain algorithm $\cdot$ Computer simulation}
\end{abstract}


\section{Introduction}
\label{sec:Introduction}

One of the greatest challenges that neuroscientists are facing today is to decode the highly complex architecture and connectivity of nerve fibers in the human brain, the so-called \textit{connectome} \cite{behrens12,sporns05,sporns09}. In recent years, the neuroimaging technique \textit{Three-dimensional Polarized Light Imaging (3D-PLI)} has proven its potential to reconstruct the spatial fiber architecture of human post-mortem brains with a resolution of a few micrometers \cite{MAxer11_1,MAxer11_2}. It enables not only to investigate the course of long-range fiber bundles but also of single fibers, which makes 3D-PLI a bridging technology between the macroscopic and the microscopic scale.

To validate the reconstructed fiber orientations, numerical simulations are used. By comparing the known underlying fiber architecture of the simulation model with the fiber orientations derived in a 3D-PLI measurement, possible misinterpretations in the fiber reconstruction process can be identified. The simulations also help to gain a better theoretical understanding of the interaction of polarized light with brain tissue and to improve the accuracy and reliability of the reconstructed fiber orientations.


\section{Three-dimensional Polarized Light Imaging (3D-PLI)}
\label{sec:3D-PLI}

The measurement and signal analysis of 3D-PLI have been described in detail by Axer et al.\ \cite{MAxer11_1,MAxer11_2}.
Here, we describe only the basic principles that are needed for the presented simulation approaches. 
\subsection{Measurement}
Post-mortem brains are fixated, frozen, and cut with a cryotome into histological sections with a thickness of about $d=70\,\upmu$m. The brain sections are embedded in a glycerin solution and placed in a polarimeter that measures the birefringence (optical anisotropy) of the brain tissue. Part of the birefringence arises from the highly ordered arrangement of lipid molecules in the myelin sheath -- an insulating layer which surrounds most of the axons in white matter \cite{goethlin13,bear71,quarles06}.
The polarimeter consists of a pair of crossed linear polarizers and a quarter-wave retarder which are rotated by angles $\rho \in$ \{0$^{\circ}$, 10$^{\circ}$, $\dots$, 170$^{\circ}$\} around the stationary brain section (see Fig.\ \ref{fig:3D-PLI_MatrixCalculus}a). The setup is illuminated by a light source with wavelength $\lambda = 525\,$nm and the transmitted light intensity is recorded by a CCD camera for each rotation angle.
\subsection{Signal Analysis}
\begin{figure}
\centering
\fbox{\includegraphics[width=\textwidth]{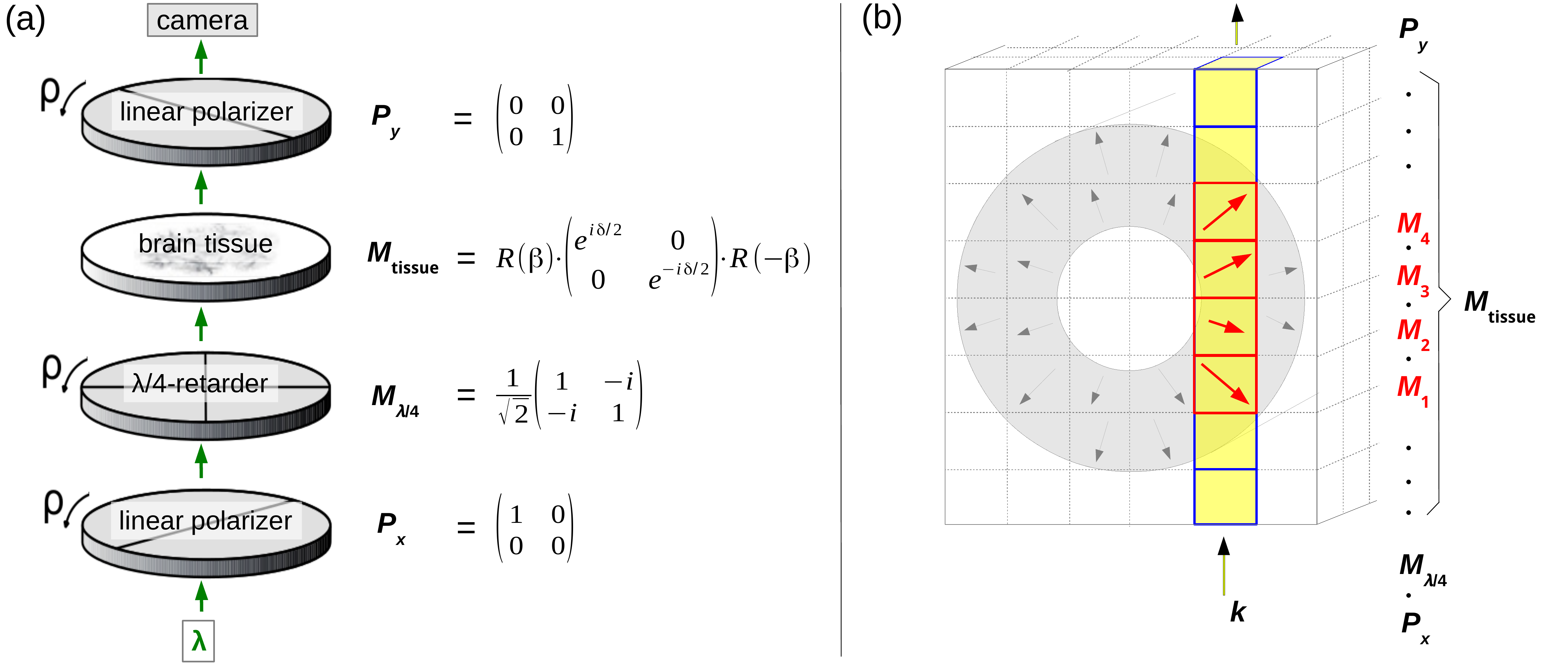}}
\caption{(a) Experimental setup of 3D-PLI and associated Jones matrices of the optical elements (b) Simulation of 3D-PLI by means of the Jones matrix calculus illustrated for a large fiber: Each myelin voxel (gray) is represented by the Jones matrix of an optical retarder ($M_j$) whose axis is oriented in direction of the optic axis (arrows). All Jones matrices along the optical path of one image pixel (highlighted column) are multiplied. (Source: Menzel et al.\ \cite{menzel15_OSA})}
\label{fig:3D-PLI_MatrixCalculus}
\end{figure}

\paragraph{\emph{\textbf{Jones Matrix Calculus.}}} 
For the analysis of the resulting light intensity profile $I(\rho)$, the \textit{Jones matrix calculus} is used \cite{jones41,jones42}: Each optical element of the polarimeter is represented by a $2 \times 2$ matrix (Jones matrix) and the electric field vector of the outgoing light $\vec{E}$ is computed by multiplying the associated Jones matrices:
\begin{align}
	\vec{E} = P_y \cdot M_{\text{tissue}} \cdot M_{\lambda/4} \cdot P_x \cdot \vec{E}_0\,.
	\label{eq:E}
\end{align}
Here, $\vec{E}_0$ represents the electric field vector of the incident light. $P_x$, $P_y$, and $M_{\lambda/4}$ are the Jones matrices of the linear polarizers and the quarter-wave retarder, respectively (see Fig.\ \ref{fig:3D-PLI_MatrixCalculus}a for definition). The birefringent brain tissue is represented by the Jones matrix of an optical retarder ($M_{\text{tissue}}$) that introduces a phase shift $\delta$ between the polarization component along the retarder axis and the polarization component perpendicular to it. The retarder axis (optic axis) is considered to be oriented in direction of the nerve fibers (with in-plane direction angle $\phi$ and out-of-plane inclination angle $\alpha$, in the following referred to as \textit{direction} and \textit{inclination}). Relative to the axis of the rotating polarizers, the retarder axis describes an in-plane rotation with rotation angle $\beta = \phi - \rho$:
\begin{align}
	M_{\text{tissue}}
		&= R(\beta) \cdot M_{\delta} \cdot R(-\beta) \notag \\
		&=	\begin{pmatrix} \cos\beta  & -\sin\beta \\ 
                			\sin\beta  & \cos\beta
    		\end{pmatrix} \,
			\begin{pmatrix} e^{\operatorname{i} \delta/2} &  0 			\\ 
                			0 			 &  e^{-\operatorname{i} \delta/2}
    		\end{pmatrix}
    		\begin{pmatrix} \cos\beta  & \sin\beta \\ 
                			-\sin\beta & \cos\beta
    \end{pmatrix} \,,
    \label{eq:M_tissue} \\
    \notag \\
    \delta &\approx \frac{2\pi}{\lambda}\, d \, \Updelta n \, \cos^2\alpha \,,
    \label{eq:delta}
\end{align}
with $\lambda$ being the wavelength of the light source, $d$ the thickness of the measured brain section, and $\Updelta n$ the local birefringence of the brain tissue \cite{MAxer11_1,MAxer11_2,menzel15}. 

The transmitted light intensity per pixel can be computed using $I_{\text{theo}} \propto \vert \vec{E} \vert^2$ and Eqs.\ (\ref{eq:E}) and (\ref{eq:M_tissue}):
\begin{align}
	I_{\text{theo}}(\rho) = \frac{I_T}{2}\left(1 + \sin\Big(2(\rho - \phi)\Big)\,\sin\delta \right)\,.
	\label{eq:I_theo}
\end{align}
Here, $I_T \propto \vert \vec{E}_0 \vert^2$ is twice the average transmitted light intensity per pixel (in the following referred to as \textit{transmittance}) and $\vert\sin\delta\vert$ the \textit{retardation} per pixel.

\paragraph{\emph{\textbf{Fourier Analysis.}}}
To derive the spatial fiber orientation ($\phi$, $\alpha$) for each image pixel, the measured intensity profile $I(\rho)$ is analyzed by means of a discrete harmonic Fourier analysis. 

Every set of $N$ data points can be represented by a Fourier series with at most $N$ coefficients ($N/2^{\text{th}}$ order):
\begin{align}
	I(\rho) &= a_{0} + \sum_{n=1}^{N/2} \Big(a_{n} \cos(n \rho) + b_{n} \sin(n \rho) \Big),
	\label{eq:FourierSeries} \\
	a_{0} = \frac{1}{N} \sum_{i=1}^N I(\rho_i)\, , \,\,\,
	a_{n} &= \frac{2}{N} \sum_{i=1}^N I(\rho_i) \cos(n \rho_i)\, , \,\,\,
	b_{n} = \frac{2}{N} \sum_{i=1}^N I(\rho_i) \sin(n \rho_i)\,.
	\label{eq:ExpFourierCoefficients}
\end{align}

Using $\sin(x-y) = \sin x \, \cos y - \cos x \, \sin y$, Eq.\ (\ref{eq:I_theo}) can be written in terms of a Fourier series with Fourier coefficients of zeroth and second order \cite{MAxer11_1,glazer96}:
\begin{align}
I_{\text{theo}}(\rho) &= \frac{I_T}{2} + \frac{I_T}{2}\,\sin\delta\,\cos(2\phi)\,\sin(2\rho)
- \frac{I_T}{2}\,\sin\delta\,\sin(2\phi)\,\cos(2\rho) \\
&\equiv a'_0 + a'_2\,\cos(2\rho) + b'_2\,\sin(2\rho)\,,
\label{eq:I_Fourier} \\
a'_0 &= \frac{I_T}{2}\, , \,\,\,
a'_2 = -\frac{I_T}{2}\,\sin\delta\,\sin(2\phi) \, , \,\,\,
b'_2 = \frac{I_T}{2}\,\sin\delta\,\cos(2\phi)\,.
\label{eq:FourierCoefficients}
\end{align}

To determine the transmittance $I_T$, the direction angle $\phi$, and the retardation $\lvert \sin\delta \rvert$ from the light intensities $I(\rho_i)$ measured at rotation angles $\rho_i \in \{0, 10^{\circ}, ..., 170^{\circ}\}$ , we assume $a_0 = a'_0$, $a_2 = a'_2$, $b_2 = b'_2$, and $b_4 = b'_4$, whereby the Fourier coefficients $a_0$, $a_2$, and $b_2$ are computed using Eq.\ (\ref{eq:ExpFourierCoefficients}), with $n=2$ and $N=18$. By rearranging Eq.\ (\ref{eq:FourierCoefficients}), we obtain: 
\begin{align}
I_T &= 2\,a_0 \,, \label{eq:I_0_transmittance}\\
\phi &= \frac{{\rm atan2} (-a_2, b_2)}{2} \,, \label{eq:phi_direction} \\
\vert \sin\delta \vert &= \frac{\sqrt{a_2^2 + b_2^2}}{a_0} \,, \label{eq:r_retardation}
\end{align}
where ${\rm atan2}$ is the arctangent with two arguments.\footnote{The function ${\rm atan2}(x,y)$ denotes the angle (in radians) between the positive x-axis and the point $(x,y)$. The angle is positive for $y>0$ and negative for $y<0$.}
The inclination angle $\alpha$ can be calculated from the retardation $\vert \sin\delta \vert$ by rearranging Eq.\ (\ref{eq:delta}).

The computed fiber orientations ($\phi$, $\alpha$) of the measured brain section are visualized in a so-called \textit{fiber orientation map (FOM)} (cf.\ Fig.\ \ref{fig:OpticChiasm}).
\newpage


\section{Simulation of 3D-PLI by means of the Jones Matrix Calculus}
\label{sec:JonesMatrixCalculus}

One possibility to simulate the interaction of polarized light with brain tissue is by using the Jones matrix calculus. Instead of representing the whole brain tissue (per pixel) by a single retarder matrix (as in Eq.\ (\ref{eq:M_tissue})), the birefringence of the myelin sheaths is modeled by multiple optical retarder elements (Jones matrices). For more details, see Menzel et al.\ \cite{menzel15} and Dohmen et al.\ \cite{dohmen15}.


\subsection{Simulation Method}
For the simulation, the nerve fibers are replaced by hollow tubes representing the surrounding myelin sheaths. The simulation volume is discretized into small cubic volume elements (voxels, indicated by the gray mesh in Fig.\ \ref{fig:3D-PLI_MatrixCalculus}b) and each myelin voxel is represented by the Jones matrix of an optical retarder with the retarder axis oriented along the optic axis of the myelin sheath (indicated by the arrows in Fig.\ \ref{fig:3D-PLI_MatrixCalculus}b). 

To generate a synthetic 3D-PLI image series, a modified version of the Jones matrix calculus described in Sec.\ \ref{sec:3D-PLI} is used whereby $M_{\text{tissue}}$ in Eq.\ (\ref{eq:M_tissue}) is replaced by the product of $N$ retarder matrices that represent the myelin voxels along the optical path of one image pixel (indicated by the highlighted column in Fig.\ \ref{fig:3D-PLI_MatrixCalculus}b):
\begin{align}
	\vec{E} = P_y \cdot (M_{N}\cdot M_{N-1} \cdots M_1) \cdot M_{\lambda /4} \cdot P_x \cdot \vec{E}_0\,.
	\label{eq:E_}
\end{align}
The synthetic 3D-PLI image series is interpreted by applying the same Fourier analysis as for the experimental data (see Sec.\ \ref{sec:3D-PLI}). The generated FOM can directly be compared to experimental results.


\subsection{Results}
\begin{figure}
\centering
\includegraphics[width=\textwidth]{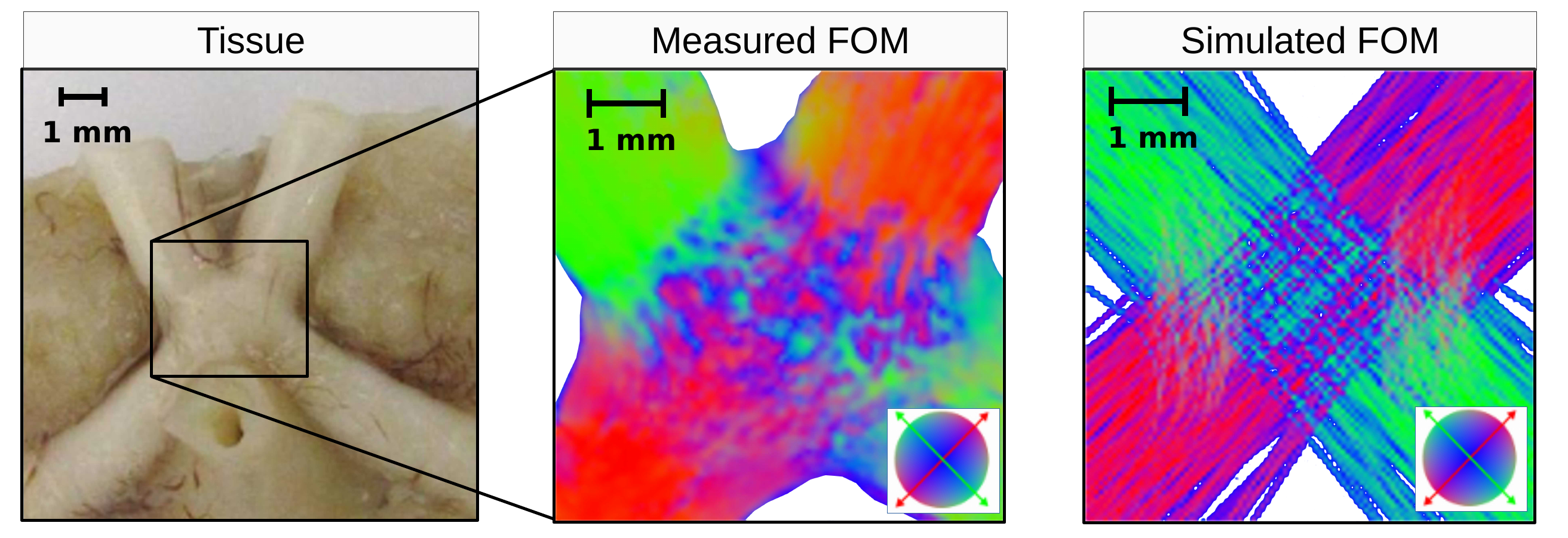}
\caption{Measured and simulated FOMs of the optic chiasm of a hooded seal, adapted from Dohmen et al.\ \cite{dohmen15}}
\label{fig:OpticChiasm}
\end{figure}

A comparison of a measured and a simulated FOM of the optic chiasm of a hooded seal (see Fig.\ \ref{fig:OpticChiasm}) demonstrates that the simulation approach based on the simple Jones matrix calculus can be used to make hypotheses on the underlying fiber structure \cite{dohmen15}. Even though the employed model of crossing and non-crossing fibers is quite simple, the most dominant features of the measured FOM are reproduced.\\

\section{Simulation of 3D-PLI by means of a 3D Maxwell Solver}
\label{sec:MaxwellSolver}

Although the previous simulation approach is already quite successful in reproducing 3D-PLI measurements, it is limited by the assumptions made in the Jones matrix calculus and the fact that only the molecular birefringence of the myelin sheaths is considered. To account for scattering and interference, we use a more sophisticated simulation approach: The propagation of the polarized light wave through the brain tissue is simulated by a massively parallel 3D Maxwell solver based on an unconditionally stable \textit{Finite-Difference Time-Domain (FDTD)} algorithm \cite{taflove}. 
%
%
\subsection{Simulation Method}
\begin{figure}
\centering
\fbox{\includegraphics[width=0.95 \textwidth]{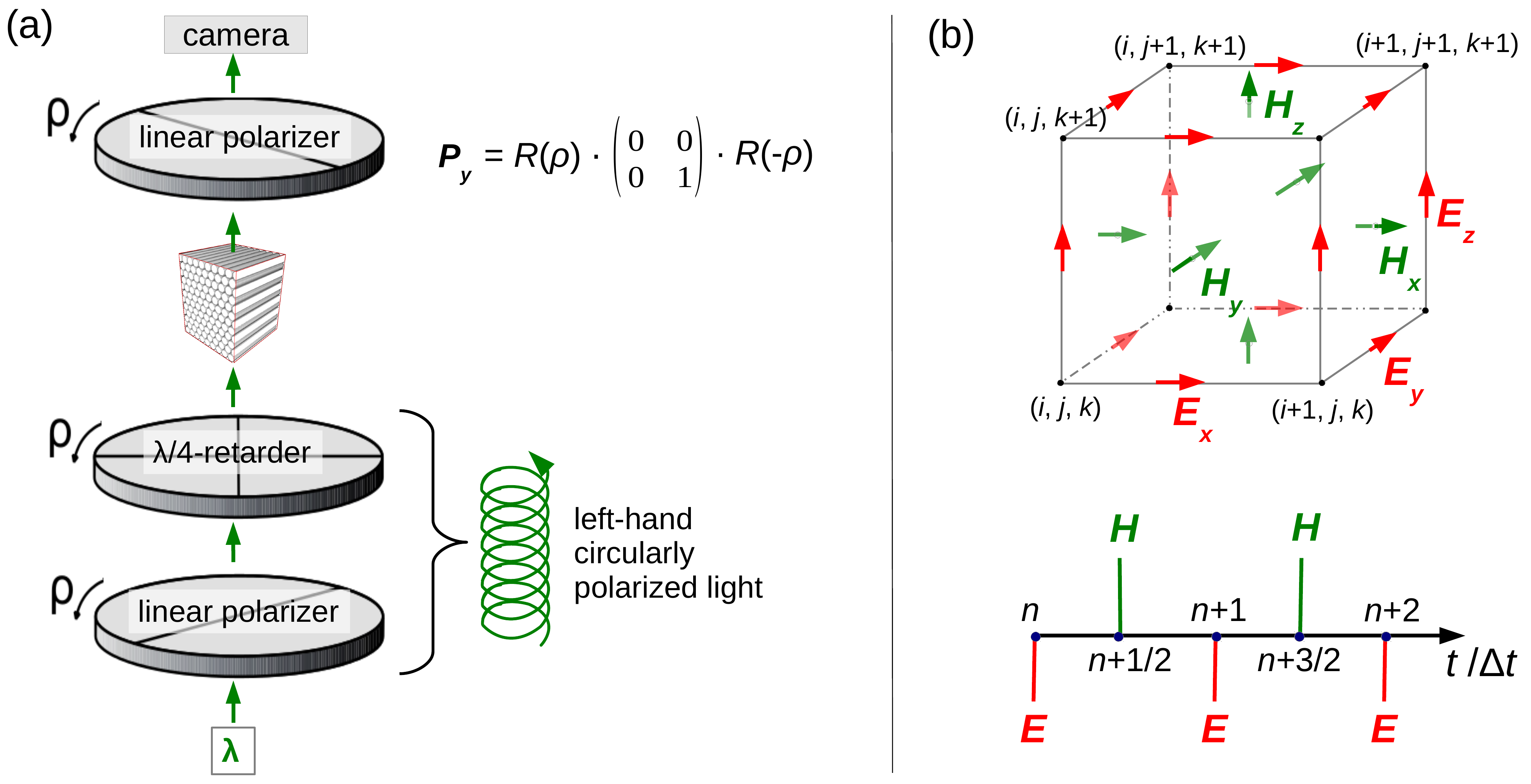}}
\caption{Simulation principles of the 3D Maxwell solver:
(a) The software TDME3D simulates the propagation of left-hand circularly polarized light through a given fiber configuration. The resulting electric field components are multiplied with the Jones matrix of a rotated linear polarizer.
(b) The upper figure shows a unit cell of the cubic Yee grid used for the discretization of space (After: Yee \cite{yee66}). The lower figure illustrates the leapfrog time-stepping scheme used for the discretization of time.}
\label{fig:MaxwellSolver}
\end{figure}

\paragraph{\emph{\textbf{Finite-Difference Time-Domain (FDTD) Algorithm.}}}
The FDTD algorithm \cite{taflove} numerically computes the components of the electromagnetic field by discretizing space and time and approximating Maxwell's curl equations by so-called \textit{finite differences}: The Maxwell equations are discretized using the Yee cell \cite{yee66}, see top panel Fig.\ \ref{fig:MaxwellSolver}b, such that each component of the electric field $\vec{E}$ is surrounded by four components of the magnetic field $\vec{H}$ and vice versa.
The propagation of the electromagnetic field in time is computed iteratively using a \textit{leapfrog time-stepping scheme} (see lower Fig.\ \ref{fig:MaxwellSolver}b): The components of the $\vec{E}$-field at a given time t are computed from the values of the $\vec{H}$-field at time $(t - \Updelta t/2)$ and from the values of the $\vec{E}$-field at time $(t - \Updelta t)$, where $\Updelta t$ is a globally defined time step. The components of the $\vec{H}$-field at time $(t + \Updelta t/2)$ are computed analogously from the values of the $\vec{E}$-field at time $t$ and from the values of the $\vec{H}$-field at time $(t - \Updelta t/2)$. 
The time-dependent electromagnetic fields are computed at every point in space using Maxwell's curl equations:
\begin{align}
\frac{\partial \vec{E}}{\partial t} &= \frac{1}{\epsilon} \Big[ \vec{\nabla} \times \vec{H} - (\vec{J}_{\text{source}} + \sigma_{e} \vec{E} ) \Big]\,, 
\label{eq:MaxwellCurl1} \\
\frac{\partial \vec{H}}{\partial t} &= - \frac{1}{\mu} \Big[ \vec{\nabla} \times \vec{E} + (\vec{M}_{\text{source}} + \sigma_{m} \vec{H} ) \Big]\,,
\label{eq:MaxwellCurl2}
\end{align}
where $\epsilon$ and $\mu$ are the electric permittivity and the magnetic permeability, $J_{\text{source}}$ and $M_{\text{source}}$ are the electric and magnetic current densities acting as independent sources of the electric and magnetic field energy, and $\sigma_{\text{e}}$ and $\sigma_{\text{m}}$ are the electric conductivity and the equivalent magnetic loss, respectively.

The spatial and temporal derivatives of the electric and magnetic fields are approximated by \textit{second-order central differences}:
\begin{align}
\frac{\partial u_{i,j,k}^n}{\partial x} &= \frac{u_{i+\frac{1}{2}, j, k}^{n} - u_{i-\frac{1}{2},j,k}^n}{\Updelta x} + O \Big[ \left(\Updelta x \right)^2 \Big]\,, 
\label{eq:FiniteDifferenceApprox1} \\
\frac{\partial u_{i,j,k}^n}{\partial t} &= \frac{u_{i , j, k}^{n+\frac{1}{2}} - u_{i,j,k}^{n-\frac{1}{2}}}{\Updelta t} + O \Big[ \left(\Updelta t \right)^2 \Big]\,,
\label{eq:FiniteDifferenceApprox2}
\end{align}
where $u_{i,j,k}^n$ represents the electric and magnetic fields evaluated at a discrete point in space ($i \Updelta x$, $j \Updelta y$, $k \Updelta z$) and a discrete point in time ($n \Updelta t$). This approximation allows to interleave the electric and magnetic field components in space and time at intervals of $\Updelta x/2$ and $\Updelta t/2$ and thus to implement the leapfrog time-stepping algorithm. 

\paragraph{\emph{\textbf{Maxwell Solver Software.}}}
For the simulations, we use the software \textit{TDME3D} $-$ a massively parallel 3D Maxwell solver that is based on an unconditionally stable FDTD algorithm. The algorithm makes use of the formal solution of Maxwell's equations in matrix form and the \textit{Lie-Trotter-Suzuki product formula approach}. For more details, see De Raedt \cite{deRaedt}.

The software solves Maxwell's equations for arbitrary (non-)periodic structures that are illuminated by arbitrary incident plane waves and that consist of linear, isotropic, lossy materials with known permeability, permittivity, and conductivity. The simulations are performed on the \textit{JUQUEEN} supercomputer \cite{juqueen} at the Forschungszentrum J{\"u}lich, Germany.


\paragraph{\emph{\textbf{Simulation of the Polarimetric Setup.}}}
The Maxwell solver computes the electromagnetic field behind a tissue sample from the given geometric and optical properties of the sample and the incident plane wave. In order to simulate a standard 3D-PLI measurement, the polarimetric setup needs to be taken into account (see Fig.\ \ref{fig:MaxwellSolver}a): After passing the first linear polarizer and the quarter-wave retarder, the light is left-hand circularly polarized. The propagation of this light wave through the sample is computed by TDME3D. The resulting electric field components ($E_x$, $E_y$, $E_z$) are then processed by a second linear polarizer rotated by angles $\rho$, yielding $\tilde{E}_x(\rho)$, $\tilde{E}_y(\rho)$, and $\tilde{E}_z(\rho)$. The x- and y-components of $\vec{\tilde{E}}$ are computed by multiplying $\vec{E}$ with the Jones matrix of a rotated linear polarizer ($R(\rho) \cdot P_y \cdot R(-\rho)$, cf.\ Sec.\ \ref{sec:3D-PLI}):
\begin{align}
	\begin{pmatrix} \tilde{E}_x			\\ 
					\tilde{E}_y
    \end{pmatrix}				&= \begin{pmatrix} \cos\rho & -\sin\rho			\\ 
										\sin\rho & \cos\rho
    					\end{pmatrix} \,
    					\begin{pmatrix} 0\,\, & 0			\\ 
										0\,\, & 1
    					\end{pmatrix} \,
    					\begin{pmatrix} \cos\rho & \sin\rho			\\ 
										-\sin\rho & \cos\rho
    					\end{pmatrix} \,
    					\begin{pmatrix} E_x			\\ 
										E_y
    					\end{pmatrix}    			\\	
    &= 	\begin{pmatrix} \cos\rho \big(E_x \cos\rho + E_y \sin\rho\big)			\\ 
						\sin\rho \big(E_x \cos\rho + E_y \sin\rho\big)
    	\end{pmatrix}\,.
    	\label{eq:E_x,y}
\end{align}
The z-component of $\vec{\tilde{E}}$ is computed by applying Maxwell's equation in free space:
\begin{align}
	\text{div} \vec{\tilde{E}} = 0  \,\, \Leftrightarrow \,\,
	\tilde{E}_z &= - \frac{1}{k_z} \big(k_x \tilde{E}_x + k_y \tilde{E}_y \big) \\
				&\overset{(\ref{eq:E_x,y})}{=} - \frac{k_x \cos\rho + k_y \sin\rho}{k_z} \big(E_x \cos\rho + E_y \sin\rho\big)\,,
	\label{eq:E_z}
\end{align} 
where $\vec{\tilde{E}} = \vec{\tilde{E}}_0 \, e^{i (\vec{k}\cdot\vec{r} - \omega t + \varphi)}$ (monochromatic plane wave) has been used.

The light intensity recorded by the camera is given by the absolute squared value of the electric field vector:
\begin{align}
	I \propto \vert \tilde{E}_x \vert^2 + \vert \tilde{E}_y \vert^2 + \vert \tilde{E}_z \vert^2\,.
\end{align} 

The x- and y-components of the electric field yield Fourier coefficients of zeroth and second order in $\rho$:
\begin{align}
	\vert \tilde{E}_x \vert^2 + \vert \tilde{E}_y \vert^2
	&\overset{(\ref{eq:E_x,y})}{=} \cos^2\rho \, \vert {E}_x \vert^2 + \sin^2\rho \, \vert {E}_y \vert^2   + \sin\rho \cos\rho \big( E_x E_y^{\ast} + E_x^{\ast} E_y \big) \\
	&= \frac{1}{2} \Big( \vert {E}_x \vert^2 + \vert {E}_y \vert^2 \Big)
	+ \frac{1}{2} \Big( \vert {E}_x \vert^2 - \vert {E}_y \vert^2 \Big) \cos(2\rho)  \\
	&\,\,\, + \frac{1}{2} \Big( E_x E_y^{\ast} + E_x^{\ast} E_y \Big) \sin(2\rho) \\
	&\equiv c_0 + c_2\, \cos(2\rho) + d_2\, \sin(2\rho)\,.
\end{align}

Similar analytical calculations show that the z-component of the electric field yields Fourier coefficients of zeroth, second, and fourth order in $\rho$:
\begin{align}
	\vert \tilde{E}_z \vert^2 &\overset{(\ref{eq:E_z})}{=} e_0 + e_2\, \cos(2\rho) + f_2\, \sin(2\rho)
+ e_4\, \cos(4\rho) + f_4\, \sin(4\rho)\,,
 \label{eq:E_z_Fourier}
\end{align}
where $e_n$ and $f_n$ are analytical functions of the wave vector $\vec{k}$ and $E_{x,y}$.

The transmitted light intensity $I(\rho)$ can therefore be represented by means of a Fourier series with Fourier coefficients $a_0$, $a_2$, $b_2$, $a_4$, and $b_4$:
\begin{align}
	I(\rho) &= a_0 + a_2\, \cos(2\rho) + b_2\, \sin(2\rho) + a_4\, \cos(4\rho) + b_4\, \sin(4\rho), \\
	a_0 &= c_0 + e_0, \,\,\,\,\, a_2 = c_2 + e_2, \,\,\,\,\, b_2 = d_2 + f_2, \,\,\,\,\, a_4 = e_4, \,\,\,\,\, b_4 = f_4\,.
	\label{eq:Maxwell_FourierCoefficients}
\end{align}
From the five Fourier coefficients, the light intensity profile $I(\rho)$ is derived for arbitrary rotation angles $\rho$.


\subsection{Results}

\begin{figure}
\centering
\fbox{\includegraphics[width=0.95 \textwidth]{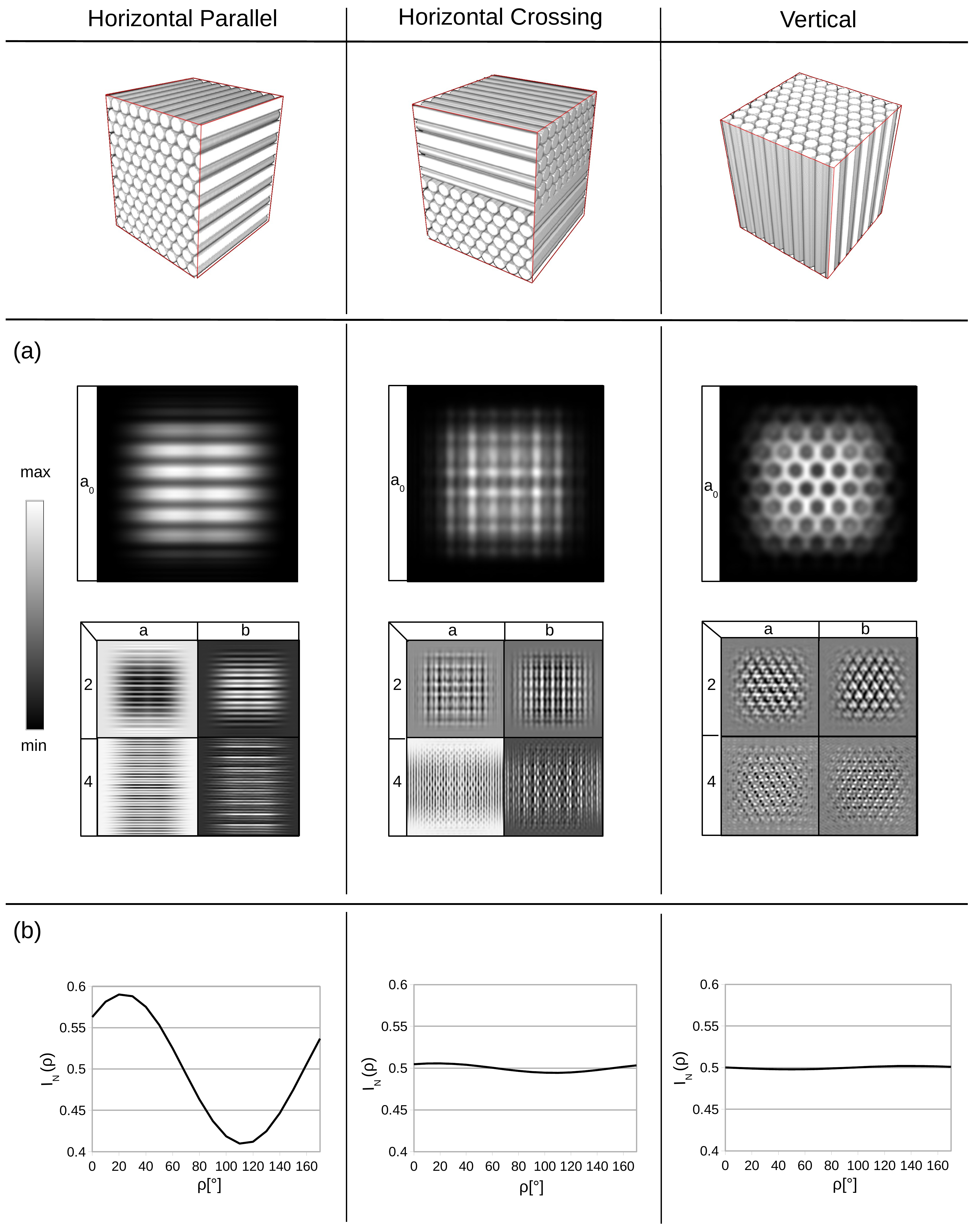}}
\caption{Simulation results of the 3D Maxwell solver computed for three samples containing horizontal parallel, horizontal crossing, and vertical fibers, respectively: (a) Fourier coefficient maps ($a_0$, $a_2$, $a_4$, $b_2$, $b_4$; cf.\ Eq.\ (\ref{eq:Maxwell_FourierCoefficients})) (b) Light intensity profiles (averaged and normalized recorded light intensity plotted against the rotation angle $\rho$)}
\label{fig:MaxwellSolver_SimResults}
\end{figure}

\paragraph{\emph{\textbf{Simulated Data.}}}
Figure \ref{fig:MaxwellSolver_SimResults} shows the computed Fourier coefficients and light intensity profiles for three samples containing horizontal parallel, horizontal crossing, and vertical fibers, respectively. 
The fibers were simulated as solid cylinders with diameters of $1\,\upmu$m and arranged in hexagonal bundles with inter-fiber distances of $0.1\,\upmu$m in a box of $10 \times 10 \times 12\,\upmu$m$^3$. The simulations were performed with uniaxial perfectly matched layer absorbing boundary conditions \cite{deRaedt07}, a Yee cell of $25\,$nm side length, and $\lambda = 525\,$nm. The refractive indices of the fibers and the surroundings were chosen as $1.47$ and $1.37$ (according to measurements of the refractive indices of myelin and the embedding glycerin solution).

Similar to a 3D-PLI measurement, the transmittance $I_T \propto a_0$ shows the underlying fiber structure (see Fig.\ \ref{fig:MaxwellSolver_SimResults}a). 
The (averaged and normalized) light intensity profiles $I(\rho)$ show a strong sinusoidal signal for horizontal parallel fibers, whereas the signal amplitude for horizontal crossing and vertical fibers is very small (see Fig.\ \ref{fig:MaxwellSolver_SimResults}b) $-$ an effect that can also be observed in a standard 3D-PLI measurement \cite{MAxer11_1,MAxer11_2,dohmen15}. This demonstrates that the Maxwell solver is able to reproduce the most dominant effects of the 3D-PLI measurement without assuming any intrinsic birefringence of the nerve fibers.

\paragraph{\emph{\textbf{Experimental Data.}}}
To derive the spatial fiber orientations in a standard 3D-PLI analysis, only the Fourier coefficients of zeroth and second order are extracted from the measured signal (see Eqs.\ (\ref{eq:I_0_transmittance})$-$(\ref{eq:r_retardation})).
However, the simulations with the Maxwell solver suggest that for non-normal incident light ($E_z \neq 0$), Fourier coefficients of fourth order will also be generated (cf.\ Eq.\ (\ref{eq:E_z_Fourier})).
 
\begin{figure}
\centering
\includegraphics[width=0.95 \textwidth]{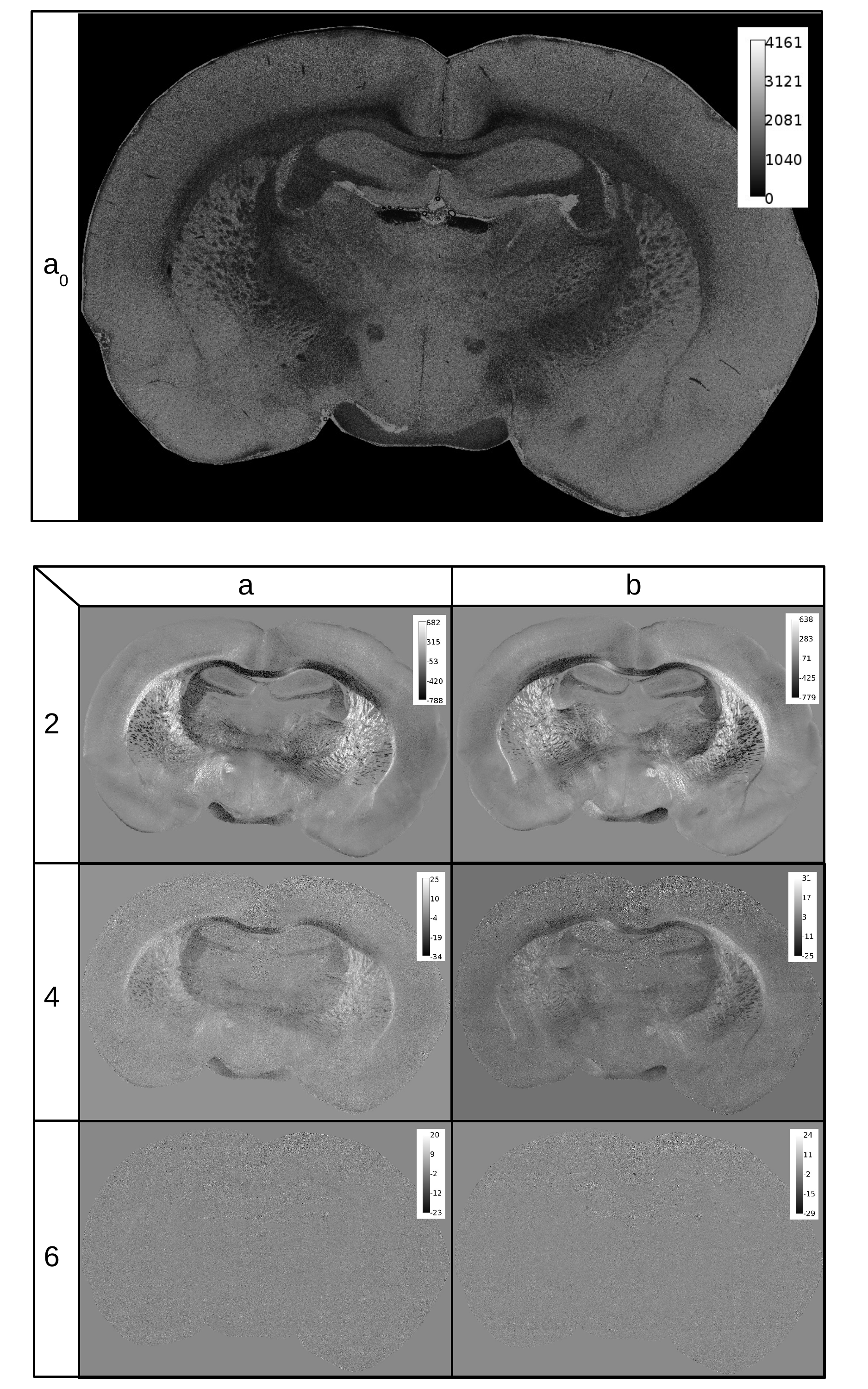}
\caption{Fourier coefficient maps ($a_0$, $a_2$, $a_4$, $a_6$, $b_2$, $b_4$, $b_6$; cf.\ Eq.\ (\ref{eq:ExpFourierCoefficients})) of a coronal rat brain section measured with 3D-PLI}
\label{fig:MaxwellSolver_ExpResults}
\end{figure}

Figure \ref{fig:MaxwellSolver_ExpResults} shows the Fourier coefficient maps (up to the sixth order) computed from a 3D-PLI measurement of a coronal rat brain section. As can be seen, the Fourier coefficients of fourth order are smaller than the Fourier coefficients of second order, but they still show the underlying tissue structure. Fourier coefficients of higher orders do not contain valuable tissue information and are probably due to noise. This suggests that non-normal incident light (e.\,g. caused by scattering) leads to Fourier coefficients of fourth order which contain valuable signal information. Therefore, $a_4$ and $b_4$ should also be taken into account when computing the fiber orientations from the measured 3D-PLI light intensity profile.


\section{Conclusion}
The 3D Maxwell solver has proven to be a valuable tool for simulating 3D-PLI. It models the interaction of polarized light with brain tissue without assuming any intrinsic birefringence of the nerve fibers. Nevertheless, the Maxwell solver reproduces the most dominant features observed in a 3D-PLI measurement and opens up new ways to improve the accuracy of the extracted fiber orientations: The FDTD simulations suggest, for example, that the Fourier coefficients of fourth order contain valuable structural information and should be incorporated in an enhanced signal analysis of 3D-PLI.


\subsubsection*{Acknowledgments.} Our work has been supported by the Helmholtz Association portfolio theme `Supercomputing and Modeling for the Human Brain', by the European Union Seventh Framework Programme (FP7/2007-2013) under grant agreement no.\ 604102 (Human Brain Project), and partially by the National Institutes of Health under grant agreement no.\ R01MH 092311.

We gratefully acknowledge the computing time granted by the JARA-HPC Vergabegremium and provided on the JARA-HPC Partition part of the supercomputer JUQUEEN \cite{juqueen} at Forschungszentrum J{\"u}lich.

We would like to thank M.\ Cremer, Ch.\ Schramm, and P.\ Nysten for the preparation of the histological brain sections.


\end{document}